# High-level Synthesis

Issam W. Damaj, Dhofar University

## Introduction

Over the years, digital electronic systems have progressed from vacuum-tube to complex integrated circuits, some of which contain millions of transistors. Electronic circuits can be separated into two groups, digital and analog circuits. Analog circuits operate on analog quantities that are continuous in value and in time, while digital circuits operate on digital quantities that are discrete in value and time (1). Examples of analog and digital systems are shown in Figure 1.

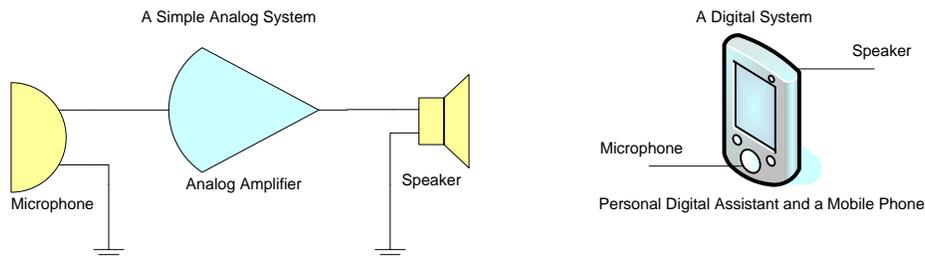

**Figure 1. A simple analog system and a digital system; the analog signal amplifies the input signal using analog electronic components. The digital system can still include analog components like a speaker and a microphone, the internal processing is digital.**

Digital electronic systems (technically referred to as digital logic systems) represent information in digits. The digits used in digital systems are the *0* and *1* that belong to the *binary* mathematical number system. In logic, the *0* and *1* values could be interpreted as *True* and *False*. In circuits, the *True* and *False* could be thought of as *High* voltage and *Low* voltage. These correspondences set the relations among logic (*True* and *False*), *binary* mathematics (*0* and *1*), and circuits (*High* and *Low*).

Logic, in its basic shape, deals with reasoning that checks the validity of a certain proposition - a proposition could be either *True* or *False*. The relation among logic, *binary* mathematics, and circuits enables a smooth transition of processes expressed in propositional logic to *binary* mathematical functions and equations (*Boolean* algebra), and to digital circuits. A great scientific wealth exist that strongly supports the relations among the three different branches of science that lead to the foundation of modern digital hardware and logic design.

*Boolean* algebra uses three basic logic operations *AND*, *OR*, and *NOT*. Truth tables and symbols of the logic operators *AND*, *OR*, and *NOT* are shown in Figure 2. Digital circuits implements the logic operations *AND*, *OR*, and *NOT* as hardware elements called "gates" that perform logic operations on binary inputs. The *AND*-gate performs an *AND* operation, an *OR*-gate performs an *OR* operation, and an *Inverter* performs the negation operation *NOT*. The actual internal circuitry of gates is built using transistors; two different circuit implementations of inverters are shown in Figure 3. Examples of *AND*, *OR*, *NOT* gates integrated circuits (*ICs* – also known as chips) are shown in Figure 4. Besides the three essential logic operations, there are four other important operations - the *NOR* (*NOT-OR*), *NAND* (*NOT-AND*), Exclusive-*OR* (*XOR*) and Exclusive-*NOR* (*XNOR*).

| Input X | Input Y | Output: X AND Y |
|---|---|---|
| False | False | False |
| False | True | False |
| True | False | False |
| True | True | True |

| Input X | Input Y | Output: X OR Y |
|---|---|---|
| False | False | False |
| False | True | True |
| True | False | True |
| True | True | True |

| Input X | Output: NOT X |
|---|---|
| False | True |
| True | False |

(a)

| Input X | Input Y | Output: X AND Y |
|---|---|---|
| 0 | 0 | 0 |
| 0 | 1 | 0 |
| 1 | 0 | 0 |
| 1 | 1 | 1 |

| Input X | Input Y | Output: X OR Y |
|---|---|---|
| 0 | 0 | 0 |
| 0 | 1 | 1 |
| 1 | 0 | 1 |
| 1 | 1 | 1 |

| Input X | Output: NOT X |
|---|---|
| 0 | 1 |
| 1 | 0 |

(b)

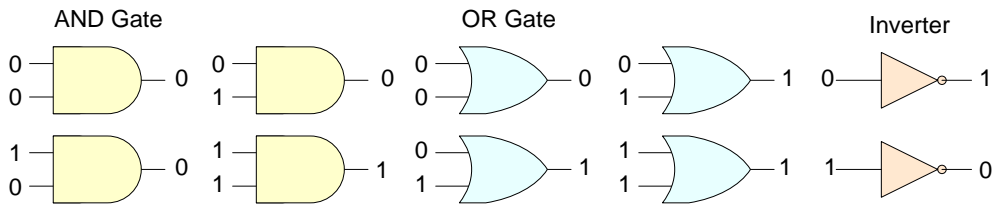

(c)

**Figure 2. (a) Truthtables for AND, OR, and Inverter. (b) Truthtables for AND, OR, and Inverter in binary numbers, (c) Symbols for AND, OR, and Inverter with their operation.**

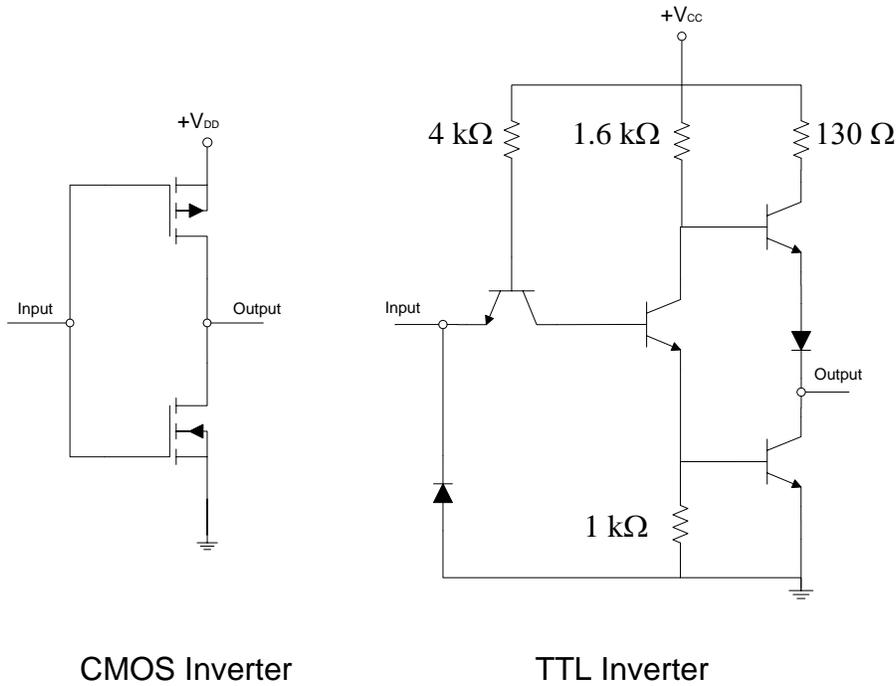

CMOS Inverter            TTL Inverter

**Figure 3. Complementary Metal-oxide Semiconductor (*CMOS*) and Transistor-Transistor Logic (*TTL*) Inverters.**

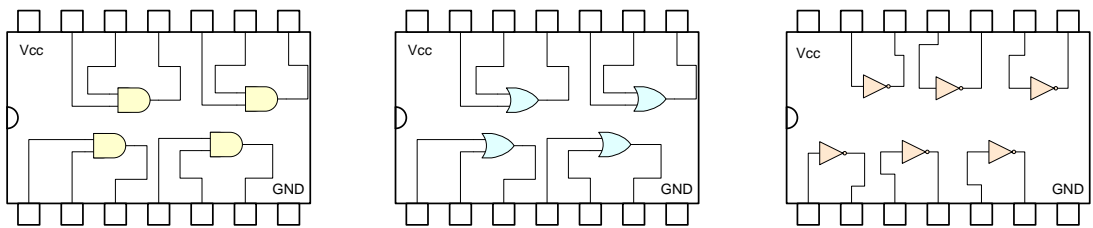

**Figure 4. The 74LS21 (*AND*), 74LS32 (*OR*), 74LS04 (*Inverter*) *TTL* ICs.**

A logic circuit is usually created by combining gates together to implement a certain logic function. A logic function could be a combination of logic variables (such as A, B, C, etc.) with logic operations; logic variables can take only the values *0* or *1*. The created circuit could be implemented using a suitable gate-structure. The design process usually starts from a specification of the intended circuit, for example, consider the design and implementation of a three variables majority function. The function *F(A, B, C)* will return a *1* (*High* or *True*) whenever the number of *1s* in the inputs is greater than or equal to the number of *0s*. The truthtable of *F* is shown in Figure 5.a. The terms that make the function *F* return a *1* are *F(0, 1, 1)*, *F(1, 0, 1)*, *F(1, 1, 0)*, or *F(1, 1, 1)*. This could be alternatively formulated as in the following equation:

$F = A'BC + AB'C + ABC' + ABC$

In Figure 5.b, the implementations using a standard *AND-OR-Inverter* gate-structure is shown. Some other specifications might require functions with more number of inputs and accordingly more complicated design process.

| Input A | Input B | Input C | Output F |
|---------|---------|---------|----------|
| 0 | 0 | 0 | 0 |
| 0 | 0 | 1 | 0 |
| 0 | 1 | 0 | 0 |
| 0 | 1 | 1 | 1 |
| 1 | 0 | 0 | 0 |
| 1 | 0 | 1 | 1 |
| 1 | 1 | 0 | 1 |
| 1 | 1 | 1 | 1 |

(a)

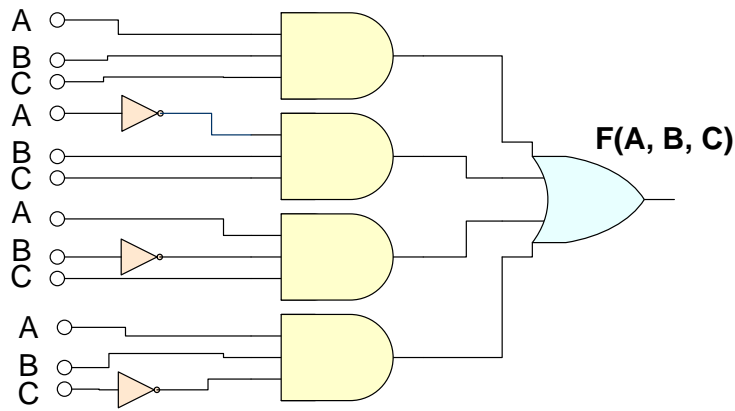

(b)
**Figure 5. (a) Truth table. (b) Standard implementation of the majority function.**

The complexity of the digital logic circuit that corresponds to a *Boolean* function is directly related to the complexity of the base algebraic function. *Boolean* functions may be simplified by several means. The simplification step is usually called optimization or minimization as it has direct effects on the cost of the implemented circuit and its performance. The optimization techniques range from simple (manual) to complex (automated using a computer).

The basic hardware design steps can be summarized in the following list:
1. Specification of the required circuit.
2. Formulation of the specification to derive algebraic equations.
3. Optimization of the obtained equations

4. Implementation of the optimized equations using suitable hardware (*IC*) technology.

The above steps are usually joined with an essential verification procedure which ensures the correctness and completeness of each design step.

Basically, there are three types of *IC* technologies that can be used to implement logic functions on (2), these are, full-custom, semi-custom, and programmable logic devices (*PLDs*). In full-custom implementations, the designer cares about the realization of the desired logic function to the deepest details including the gate-level and the transistor-level optimizations to produce a high performance implementation. In semi-custom implementations, the designer uses some ready logic-circuit blocks and completes the wiring to achieve an acceptable performance implementation in a shorter time than full-custom procedures. In *PLDs*, the logic blocks and the wiring are ready. In implementing a function on a *PLD*, the designer will only decide of which wires and blocks to use; this step is usually referred to as programming the device.

The task of manually designing hardware tends to be extremely tedious, and sometimes impossible, with the increasing complexity of modern digital circuits. Fortunately, the demand on large digital systems has been accompanied with a fast advancement in *IC* technologies. Indeed, *IC* technology has been growing faster than the ability of designers to produce hardware designs. Hence, there has been a growing interest in developing techniques and tools that facilitate the process of hardware design.

The task of making hardware design simpler has been largely inspired by the success story in facilitating the programming of traditional computers done by software designers. This success has motivated eager hardware designers to follow closely the footsteps of software designers leading to a synergy between these two disciplines creating what is called hardware/software co-design.

## Software Design

A computer is basically composed from a computational unit made out of logic components whose main task is to perform arithmetic and logic operations; this is usually called the arithmetic and logic unit (*ALU*). The computations performed by the *ALU* are usually controlled by a neighboring unit called the control unit (*CU*). The *ALU* and the *CU* construct the central processing unit (*CPU*) that is usually attached to a storage unit, memory unit, input and output units to build a typical digital computer. A simplified digital computer is shown in Figure 6.

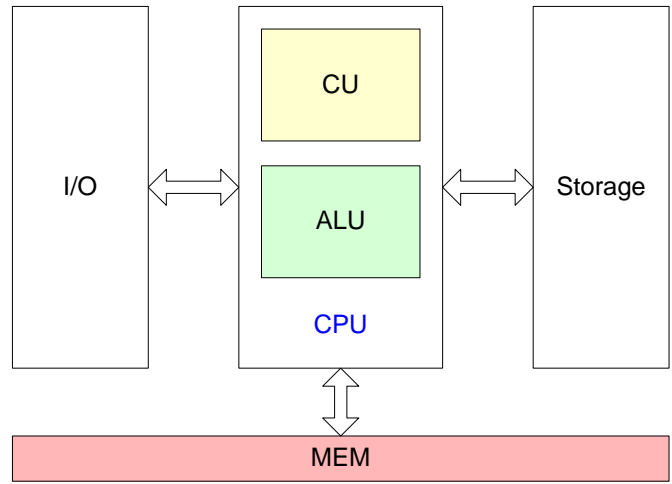

**Figure 6. A typical organization of a digital computer.**

To perform an operation using the *ALU* the computer should provide a sequence of bits (machine instruction) that include signals to enable the appropriate operation, the inputs, and the destination of the output. To run a whole program (sequence of instruction), the computations are provided sequentially to the computer. As the program sizes grow, dealing with *0s* and *1s* become difficult. Efforts for facilitating dealing with computer programs concentrated on the creation of translators that hides the complexity of dealing with programming using *0s* and *1s*. An early proposed translator produced the binary sequence of bits (machine instructions) from easy to handle instructions written using letters and numbers called assembly language instructions. The translator performing the above job is called an assembler (See Figure 7).

Before long, the limitations of assembly instructions became apparent for programs consisting of thousands of instructions. The solution came in favor of translation again; this is time the translator is called a compiler. Compilers automatically translate sequential programs, written in a high-level language like *C*, *Pascal*, etc, into equivalent assembly instructions (See Figure 7). Translators like assemblers and compilers, helped software designers ascend to higher levels of abstraction. With compilers, a software designer can code with few number of lines that are easy to understand. Then, the compiler will do the whole remaining job of translation hiding all the low-level complex details from a software designer.

## Towards Automated Hardware Design

Translation from higher levels of abstraction for software has motivated the creation of automated hardware design (synthesis) tools. The idea of hardware synthesis sounds very similar to that for software compilation. A designer can produce hardware circuits by automatically synthesizing an easy to understand description of the required circuit, provided a list of performance-related requirements.

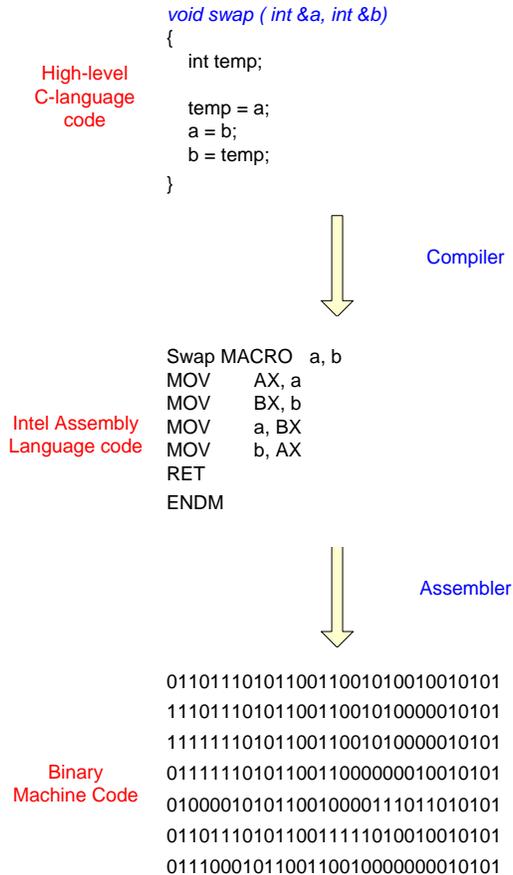

**Figure 7. The translation process of high-level programs.**

There are several advantages of automating part or all of the hardware design process and moving automation to higher levels of abstraction. Firstly, automation assures a much shorter design cycle and time to market the produced hardware. Secondly, automation allows for more exploration of different design styles since different designs can be synthesized and evaluated within a short time. Finally, with well-developed design automation tools, it may out-perform human designers in generating high quality designs.

## Hardware Design Approaches

Two different approaches emerged from the debate over ways to automate hardware design. On one hand, the capture-and-simulate proponents believe that human designers have good design experience that cannot be automated. They also believe that a designer can build a design in a bottom-up style from elementary components such as transistors and gates. Since the designer is concerned with the deepest details of the design, optimized and cheap designs could be produced. On the other hand, the describe-and-synthesis advocates believe that synthesizing algorithms can out-perform human designers. They also believe that a top-down fashion would be better suited for designing complex systems. In describe-and-synthesize methodology, the designers firstly describe the design. Then, computer aided design (*CAD*) tools can generate the physical and electrical structure. This approach describes the intended designs using special languages called hardware description languages (*HDLs*). Some *HDLs* are very similar to traditional programming languages like *C*, *Pascal*, etc. (3).

Both of these design approaches may be correct and useful at some point. For instance, circuits made from replicated small cells (like memory) are to perform efficiently if the cell is captured, simulated, and optimized to the deepest-level components (such as transistors). Another complicated heterogeneous design that will be developed and mapped onto a ready prefabricated device, like a *PLD* where no optimizations are possible on the electronics level, can be described and automatically synthesized. However, modern synthesis tools are well equipped with powerful automatic optimization tools.

## High-level Hardware Synthesis

Hardware synthesis is a general term used to refer to the processes involved in automatically generating a hardware design from its specification. High-level Synthesis (*HLS*) could be defined as the translation from a behavioral description of the intended hardware circuit into a structural description similar to the compilation of programming languages (such as, *C*, *Pascal*, etc.) into assembly language. The behavioral description represents an algorithm, equation, etc., while a structural description represents the hardware components that implement the behavioral description. In spite of the general similarity between hardware and software compilations, hardware synthesis is a multi-level and complicated task. In software compilation, you translate from a high-level language to a lower-level language, while in hardware synthesis you step through a series of levels.

To explain more on behavior, structure, and their correspondences, Figure 8 shows Gajski's Y-chart. In this chart, each axis represents a type of description (Behavioral, Structural, and Physical). On the behavioral side, the main concern is algorithms, equations, functions but no implementation. On the structural side, implementation constructs are shown; the behavior is implemented by connecting components with known behavior. On the physical side, circuit size, component placements and wire routes on the developed chip (or board) are the main focus.

The chained synthesis tasks at each level of the design process include system synthesis, register-transfer synthesis, logic synthesis, and circuit synthesis. System synthesis starts with a set of processes communicating though either shared variables or message passing. It generates a structure of processors, memories, controllers, and interface adapters from a set of system components. Each component can be described using a register-transfer language (*RTL*). *RTL* descriptions model a hardware design as circuit blocks and interconnecting wires. Each of these circuit blocks could be described using *Boolean* expressions. Logic synthesis translates *Boolean* expressions into a list of logic gates and their interconnections (netlist). The used gates could be components from a given library such as *NAND*, *NOR*, etc. In many cases, a structural description using one library must be converted into one using another library (usually referred to as technology mapping). Based on the produced netlist, circuit synthesis generates a transistor schematic from a set of input-output current, voltage and frequency characteristics or equations. The synthesized transistor schematic contains transistor types, parameters and sizes.

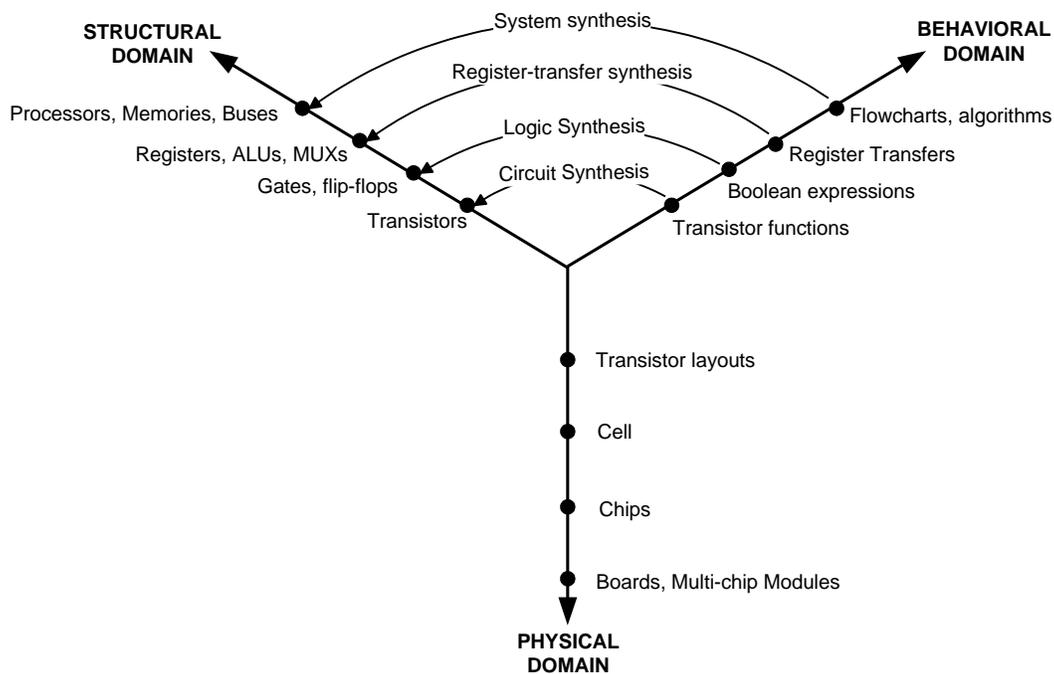

**Figure 8. Gajski's Y-chart.**

Early contributions to *HLS* were done in the 1960s. The *ALERT* (4) system was developed at *IBM*. *ALERT* automatically translates behavioral specifications written in *APL* (5) into logic-level implementations. The *MIMOLA* system (1976) generated a *CPU* from a high-level input specification (6). *HLS* witnessed a considerable growth since early 1980s, and currently plays a key role in modern hardware design.

## High-level Synthesis Tools

A typical modern hardware synthesis tool includes *HLS*, logic synthesis, placement, and routing steps as shown in Figure 9. In terms of Gajski's Y-chart vocabulary, these modern tools synthesize a behavioral description into a structural network of components. The structural network is then further synthesized, optimized, placed physically in a certain layout, and then routed through. The *HLS* step includes, firstly, allocating necessary resources for the computations needed in the provided behavioral description (Allocation stage). Secondly, the allocated resources are bind to the corresponding operations (Binding stage). Thirdly, the operations order of execution is scheduled (Scheduling stage). The output of the high-level synthesizer is an *RT*-level description. The *RT*-level description is then logically synthesized to produce an optimized netlist. Gate netlists are then converted into circuit modules by placing cells of physical elements (Transistors) into several rows and connecting input/output (*I/O*) pins through routing in the channels between the cells. The following example illustrates the *HLS* stages (Allocation, Binding, and Scheduling).

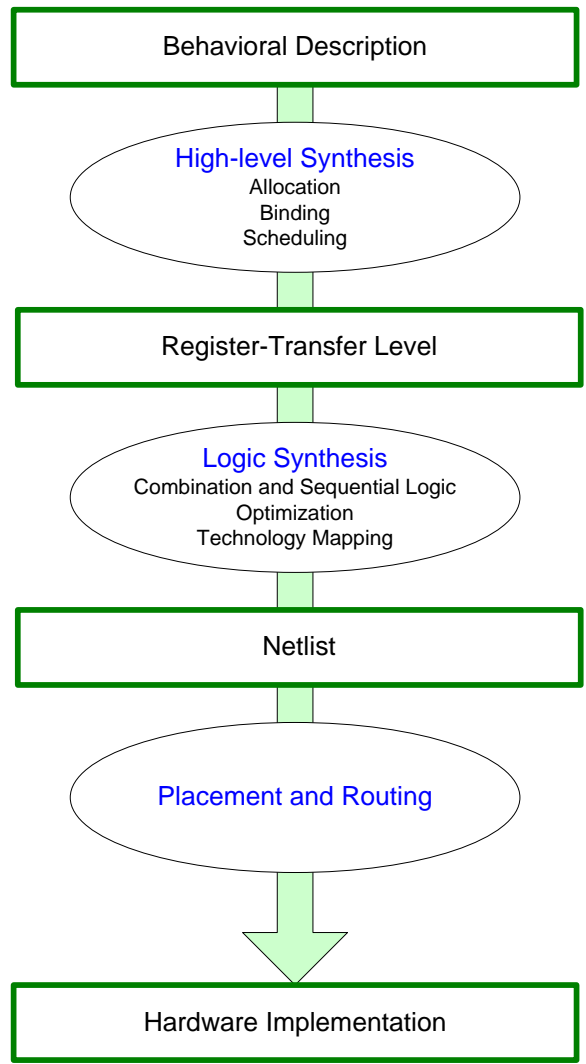

**Figure 9. The process of describe-and-synthesize for hardware development.**

Consider a behavioral specification that contains the statement, $s = a^2 + b^2 + 4b$. The variables *a* and *b* are predefined. Assume that the designer has allocated two multipliers

($m_1$ and $m_2$) and one adder ($ad$) for $s$. However, to compute $s$ a total of three multipliers and two adders could be used as shown in the dataflow graph in Figure 10.

A possible binding and schedule for the computations of $s$ is shown in Figure 11. In the first step, the multiplier $m_1$ is bind with the computation of $a^2$, and the multiplier $m_2$ is bind with the computation of $b^2$. In the second step, $m_1$ is reused to compute ($4b$); also the adder ($ad$) is used to perform ($a^2 + b^2$). In the third and last step, the adder is reused to add ($4b$) to ($a^2 + b^2$). Different bindings and schedules are possible. Bindings and schedules could be carried out to satisfy a certain optimization, for example, to minimize the number computational steps, routing, or maybe multiplexing.

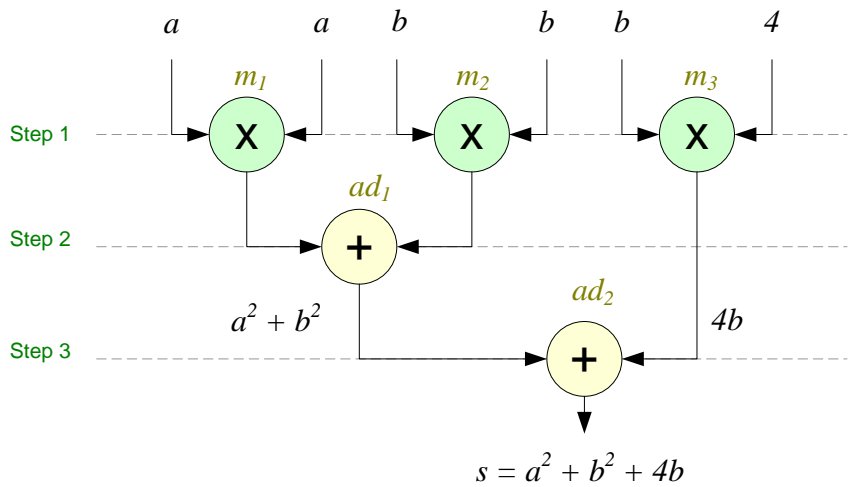

Figure 10. A possible Allocation, binding and scheduling of $s = a^2 + b^2 + 4b$.

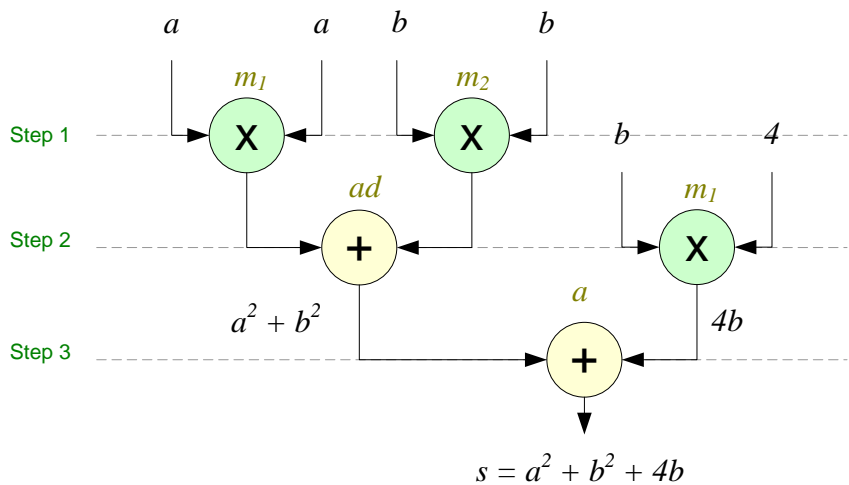

Figure 11. Another possible allocation, binding and scheduling of $s = a^2 + b^2 + 4b$.

## Hardware Description Languages

*HDLs*, like traditional programming languages, are often categorized according to their level of abstraction. Behavioral *HDLs* focus on algorithmic specifications and support constructs commonly found in high-level imperative programming languages, such as, assignment, conditionals, etc.

*Verilog* (7) and *VHDL* (Very High Speed Integrated Circuit Hardware Description Language) (8) are by far the most commonly used *HDLs* in industry. Both of these two *HDLs* support different styles for describing hardware, for example, behavioral style, structural gate-level style, etc. *VHDL* became an *IEEE* standard *1076* in 1987. *Verilog* became an *IEEE* standard *1364* in December 1995.

The *Verilog* language uses the *module* construct to declare logic blocks (with a number of inputs and outputs). In Figure 12, a *Verilog* description of a half-adder circuit is shown.

In *VHDL*, each structural block consists of an interface description and architecture. VHDL enables behavioral descriptions in Data flow and Algorithmic styles. The half-adder circuit of Figure 12 has a dataflow behavioral *VHDL* description as shown in Figure 13; a structural description is shown in Figure 14.

```
Module Half_Adder (a, b, c, s);
      input a, b;
      output c, s; //Output sum and carry.

      and Gate1 (c, a, b);    //an AND gate with two inputs a and b
                              //and one output c

      xor Gate2 (s, a, b)     //a XOR gate with two inputs a and b
                              //and one output s
endmodule
```
**Figure 12. A *Verilog* description of a half-adder circuit.**

```
entity Half_Adder is
     port (
           a: in STD_LOGIC;
           b: in STD_LOGIC;
           c: out STD_LOGIC;
           s: out STD_LOGIC);
end Half_Adder

architecture behavioral of Half_Adder is
begin

s <= (a xor b) after 5 ns;
c <= (a and b) after 5 ns;

end behavioral;
```
**Figure 13. A behavioral *VHDL* description of a Half_Adder.**

```
entity Half_Adder is
      port (
            a, b: in bit;
            c, s: out bit;);
end Half_Adder

architecture structural of Half_Adder is
      component AND2    port (x, y: in bit; o: out bit);
      component EXOR2   port (x, y: in bit; o: out bit);

begin

Gate1 : AND2 port map (a, b, c);
Gate2 : EXOR2 port map (a, b, s);

end structural;
```

**Figure 14. A structural *VHDL* description of a Half_Adder.**

Efforts for creating tools with higher levels of abstraction lead to the production of many powerful modern hardware design tools. Ian Page and Wayne Luk developed a compiler that transformed a subset of *Occam* into a netlist (9). Nearly ten years later we have seen the development of *Handel-C* (9), the first commercially available high-level language for targeting programmable logic devices (such as field programmable gate arrays - *FPGAs*).

*Handel-C* is a parallel programming language based on the theories of communicating sequential processes (*CSP*) and *Occam* with a *C*-like syntax familiar to most programmers. This language is used for describing computations which are to be compiled into hardware. A *Handel-C* program is not compiled into machine code, but into a description of gates and flip-flops, which is then used as an input to *FPGA* design software. Investments for research into rapid development of reconfigurable circuits using *Handel-C* have been largely done at *Celoxica* (11). *Handel-C* compiler comes packaged with the *Celoxica DK Design Suite*.

Almost all *ANSI-C* types are supported in *Handel-C*. Also, *Handel-C* supports all *ANSI-C* storage class specifiers and type qualifiers expect *volatile* and *register* which have no meaning in hardware. *Handel-C* offers additional types for creating hardware components such as memory, ports, buses and wires. *Handel-C* variables can only be initialized if they are global or if declared as *static or const*. Figure 15 shows *C* and *Handel-C* types and objects, in addition to the design flow of *Handel-C*. Types are not limited to width in *Handel-C* since, when targeting hardware, there is no need to be tied to a certain width. Variables can be of different widths, thus minimizing the hardware usage. For instance, if we have a variable *a* that can hold a value between 1 and 5, then it is enough to use 3 bits only (declared as: *int 3 a*).

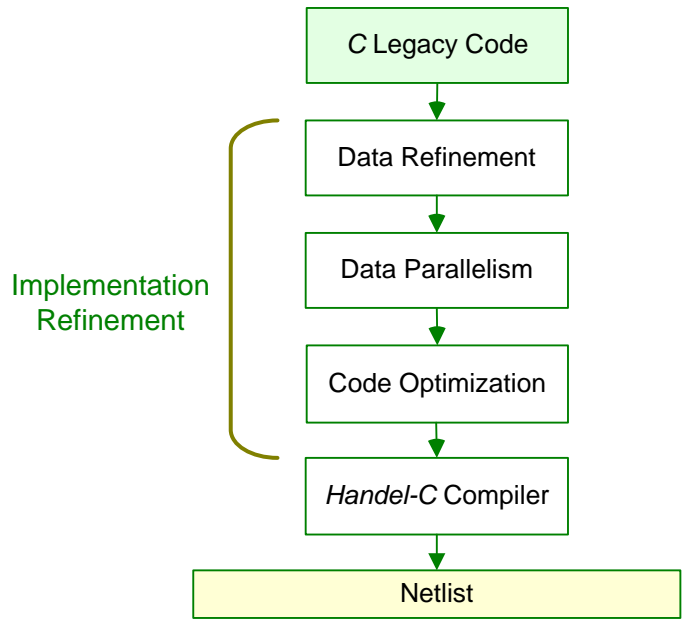

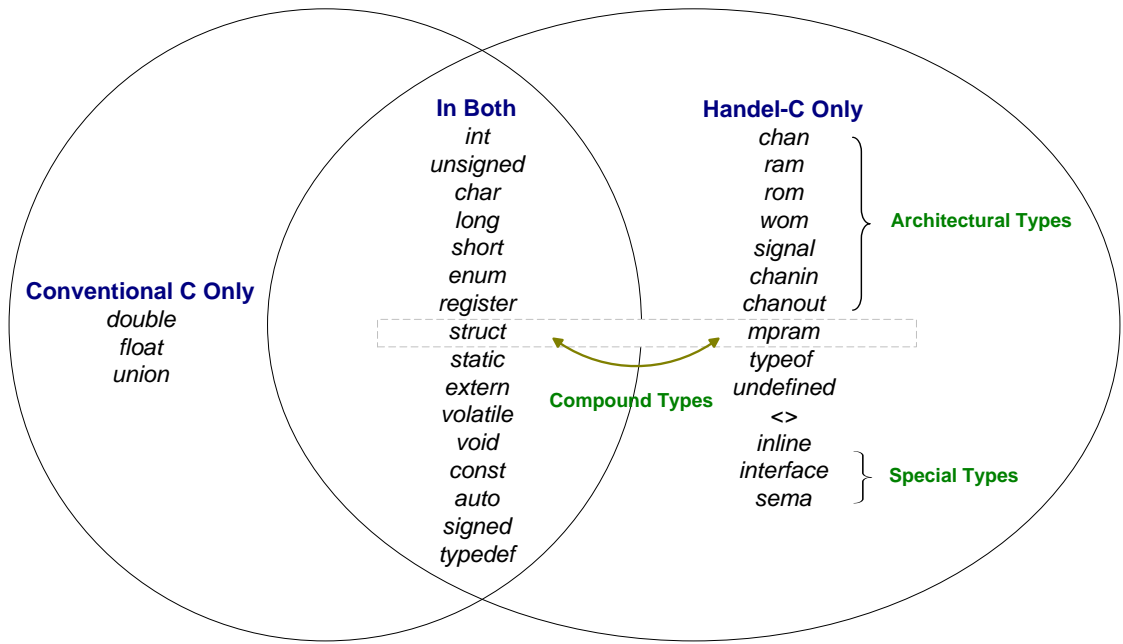

**Figure 15**. *C* and *Handel-C* types and objects. *Handel-C* types can be classified as common logic types, architectural types, compound types and special types

The notion of time in *Handel-C* is fundamental. Each assignment happens in exactly one clock cycle, everything else is "free". An essential feature in *Handel-C* is the *par* construct which executes instructions in parallel. Figure 16 provides an example showing the effect of using *par*.

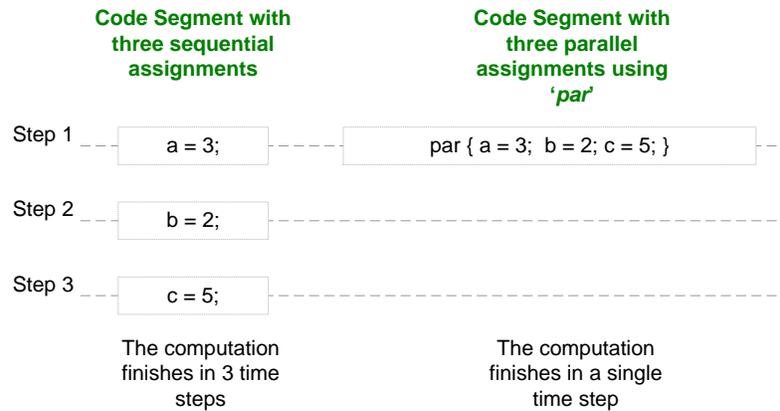

**Figure 16. Parallel execution using *par* statement**

Building on the work carried out in Oxford's Hardware Compilation Group by Page and Luk, Saul at Oxford's Programming Research Group introduced a different co-design compiler, *Dash FPGA*-Based Systems (12). This compiler provides a co-synthesis and co-simulation environment for mixed *FPGA* and processor architectures. It compiles a *C*-like description to a solution containing both processors and custom hardware.

Luk and McKeever in (13) introduced *Pebble*, a simple language designed to improve the productivity and effectiveness of hardware design. This language improves productivity by adopting reusable word-level and bit-level descriptions which can be customized by different parameter values, such as design size and the number of pipeline stages. Such descriptions can be compiled without flattening into various *VHDL* dialects. *Pebble* improves design effectiveness by supporting optional constraint descriptions, such as placement attributes, at various levels of abstraction; it also supports runtime reconfigurable designs.

Todman and Luk in (14) proposed a method that combines declarative and imperative hardware descriptions. They investigated the use of *Cobble* language, which allows abstractions to be done in an imperative setting. Designs done in *Cobble* are to benefit from efficient bit-level implementations developed in *Pebble*. Transformations are suggested to allow the declarative *Pebble* blocks to be used in *Cobbles'* imperative programs.

Weinhardt in (15) proposes a high-level language programming approach for reconfigurable computers. This automatically partitions the design between hardware and software, and synthesizes pipelined circuits from parallel *for* loops.

W. Najjar et al in (16) presented a high-level, algorithmic language and optimizing compiler for the development of image processing applications on *RC*-systems. *SA-C*, a single assignment variant of the *C* programming language, was designed for this purpose.

A prototype *HDL* called Lava is developed by Satnam Singh at *Xilinx* and Mary Sheeran and Koen Claessen at Chalmers University in Sweden (17). Lava allows circuit tiles to be composed using powerful higher-order combinators. This language is embedded in the *Haskell* lazy functional programming language. *Xilinx* implementation of *Lava* is designed to support the rapid representation, implementation and analysis of high performance *FPGA* circuits.

Besides the above advances in the area of high-level hardware synthesis, the current market has other tools employed to aid programmable hardware implementations. These tools include *Forge* compiler from *Xilinx*, *SystemC* language, *Nimble* compiler for *Agileware* architecture from *Nimbel Technology*, and *Superlog*.

*Forge* is a tool for developing reconfigurable hardware, mainly *FPGAs*. *Forge* uses *Java* with no changes to syntax. It also requires no hardware design skills. The *Forge design suite* compiles into *Verilog*, which is suitable for integration with standard *HLS* and simulation tools.

*SystemC* is based on a methodology that can be effectively used to create a cycle-accurate model of a system consisting of software, hardware and their interfaces in *C++*. *SystemC* is easy to learn for people who already use *C/C++*. *SystemC* produces an executable specification, while inconsistencies and errors are avoided. The executable specification helps to validate the system functionality before it is implemented. The momentum in building *SystemC* language and modeling platform is to find a proper solution for representing functionality, communication, and software and hardware implementations at various levels of abstraction.

The *Nimble* compiler is an *ANSI-C* based compiler for a particular type of architecture called the *Agileware*. The *Agileware* architecture consists of a general purpose *CPU* and a dynamically configurable data path coprocessor with a memory hierarchy. It can parallelize and compile the code into hardware and software without user intervention. *Nimble* can extract computationally intensive loops, turn them into data flow graphs and compile them into a reconfigurable datapath.

*Superlog* is an advanced version of *Verilog*. It adds more abstract features to the language allowing designers to handle large and complex chip designs without getting too much into details. Besides, *Superlog* adds many object oriented features as well as advanced programming construct to *Verilog*.

Other famous *HLS* and hardware design tools include *Altera's Quartus*, *Xilinx ISE*, *Mentor Graphics HDL Designer*, *Leonardo Spectrum*, *Precision Synthesis*, and *ModelSim*.

**Higher-level Hardware Design Methodologies**

The area for deriving hardware implementations from high-level specifications has been witnessing a continuous growth. The aims have been always to reach higher-levels of abstraction through correct well defined refinement steps. Many frameworks for developing correct hardware have been brought out in the literature (18, 19, 20).

Hoare et al in the Provably Correct Systems project (*ProCoS*), suggested a mathematical basis for the development of embedded and real-time computer systems. They used *FPGAs* as back-end hardware for realizing their developed designs. The framework included novel specification languages and verification techniques for four levels of development:
- Requirements definition and design.
- Program specifications and their transformation to parallel programs.
- Compilation of programs to hardware.
- Compilation of real-time programs to conventional processors.

Aiming for a short and precise specification of requirements, *ProCoS* has investigated a real-time logic to formalize dynamic systems properties. This logic provides a calculus to verify a specification of a control strategy based on finite state machines (*FSM*). The specification language *SL* is used to specify program components, and to support transformation to an *Occam*-like programming language *PL*. These programs are then transformed to hardware or machine code. A prototype compiler in *SML* has been produced, which converts a *PL*-like language to a netlist suitable for placement and routing for *FPGAs* from *Xilinx*.

Abdallah et al. (19), at London South Bank University, created a step-wise refinement approach to the development of correct hardware circuits from formal specifications. A functional programming notation is used for specifying algorithms and for reasoning about them. The specifications are realized through the use of a combination of function decomposition strategies, data refinement techniques, and off-the-shelf refinements based upon higher-order functions. The off-the-shelf refinements are inspired by the operators of *CSP* and map easily to programs in *Handel-C*. The *Handel-C* descriptions are then directly compiled into hardware.

The development of hardware solutions for complex applications is no more a complicated task with the emergence of various *HLS* tools. Many areas of application have benefited from the modern advances in hardware design, such as, automotive and aerospace industries, computer graphics, signal and image processing, security, complex simulations like molecular modeling, *DNA* matching, etc.

The field of *HLS* is continuing its rapid growth facilitating the creation of hardware and blurring more and more the border separating the processes of designing hardware and software.

## Reading List

T. Floyd , Digital Fundamentals with PLD Programming, New Jersey: Prentice Hall, 2006.

M. Mano et al., Logic and Computer Design Fundamentals, New Jersey: Prentice Hall, 2004.

F. Vahid et al., Embedded System Design: A Unified Hardware/Software Introduction, New York: John Wiley & Sons, 2002.

S. Hachtel, Logic Synthesis and Verification Algorithms, Norwell: Kluwer, 1996.

## Cross-references

Programmable Logic Devices, See *PLDs*.